\documentclass[useAMS,usenatbib]{mn2e}
\usepackage{natbib}
\usepackage{times}
\usepackage{graphicx,aas_macros}
\usepackage{hyperref}
\usepackage{amssymb}
\usepackage{ulem}
\graphicspath{{plots/}}
\def \aap {A\&A}

\def \apjl {ApJL}

\title[Modified pulsar current analysis]{Modified pulsar
current analysis: probing magnetic field evolution }
\author[A. P. Igoshev \& S.B. Popov]{A.P. Igoshev$^{1}$\thanks{E-mail:
ignotur@gmail.com;}, S.B. Popov$^{2}$\\
$^{1}$ Department of Astrophysics/IMAPP Radboud University Nijmegen P.O. Box 9010 6500 GL Nijmegen The Netherlands \\
$^{2}$ Sternberg Astronomical Institute, Lomonosov Moscow State University, Universitetsky prospekt 13, 119991, Moscow, Russia}
\begin{document}

\date{Accepted --- Received ---}


\maketitle

\label{firstpage}

\begin{abstract}
We use a modified pulsar current analysis to study magnetic field decay in
radio pulsars.  In our approach we analyse the flow, not along the
spin period axis as has been performed  in previous studies, but study the flow
along the direction of growing characteristic age, $\tau=P/(2\dot P)$.
We perform extensive tests of the method and find that in most of the cases it is
able to uncover
non-negligible magnetic field decay (more than a few tens of per cent during the
studied range of ages) in normal radio pulsars for realistic initial properties of neutron
stars.  However, precise determination of the magnetic field decay
timescale is not possible at present.  
The estimated timescale may differ by a factor of few for
different sets of initial distributions of neutron star parameters.
In addition,  
some combinations of initial distributions 
and/or selection effects can also mimic enhanced field decay. 
We apply our method to the observed sample of radio pulsars at distances
$<10$~kpc in the range 
of characteristic ages $8 \times 10^4 < \tau < 10^6$~years where, according
to our study, selection effects are minimized. 
By analysing pulsars in the Parkes Multibeam and Swinburne surveys we find that, in this range,
the field decays roughly by a factor of two. With an exponential fit this
corresponds to the decay time scale $\sim 4 \times 10^5$~yrs.
With larger statistics and better knowledge of the initial distribution of spin
periods and magnetic field strength, this method can be a powerful tool to probe 
magnetic field decay in neutron stars.
\end{abstract}


\section{Introduction} 
The pulsar current analysis is a known method to
study the evolution of radio pulsars.  It was originally proposed and
applied by \cite{pulsar_curr} and \cite{pb1981}, and more recently revised by
\cite{pulsar_curr_rev}.  It is assumed that pulsars are born in a certain
region (or regions) in the spin period~---~period derivative ($P$~--~$\dot
P$) plane, and then they move along evolutionary tracks (which depend on the
magnetic field evolution model), until they finally disappear in another part
of the $P$~--~$\dot P$ diagram.  The classical pulsar current evolves
according to a kinetic equation with a source term (see
\citealt{pulsar_curr_rev} for details).  One of the main results of this
technique is an estimate of a total birthrate.  In addition,
information about initial spin period distribution can be uncovered
by this method.  For example, this technique provided evidence in favour of
so-called ``injection'' in the pulsar current at $P\sim 0.5$ s
(\citealt{pulsar_curr}, although this result has been questioned in later
studies, see \citealt{pulsar_curr_rev}).

\begin{figure}
\center{\includegraphics[width=84mm]{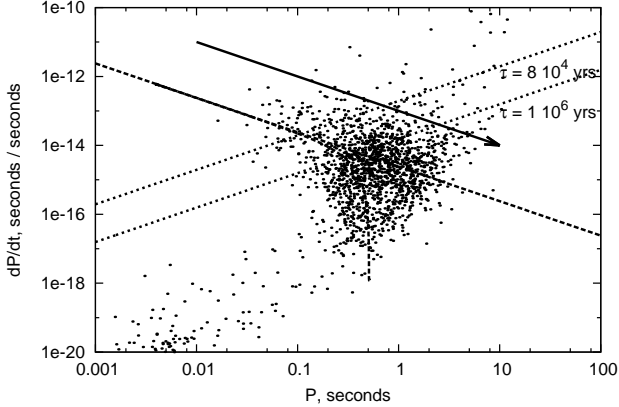}}
\caption{$P$~---~$\dot P$ diagram. Black dots represent normal pulsars from the ATNF catalogue. 
With dashed lines we show two evolutionary tracks: with constant  and
with an exponentially decaying field. Dotted lines correspond to two values of
$\tau$. Finally, the solid arrow (corresponding to the gradient of
characteristic age)
 illustrates the direction in which $\tau$ is growing. In our approach we
study pulsar current along this direction.
}
\label{P_dotP}
\end{figure}

In this article we propose a modification to the pulsar current analysis. 
The main difference from the standard technique is that we look at the
pulsar current along the spin-down age, $\tau$,  direction (black arrow in
Fig.~\ref{P_dotP}) 
instead of the spin period axis. This approach has an advantage with respect
to the standard pulsar current analysis: if the magnetic field of a neutron
star rapidly decays, then the spin period grows very slowly. However, the
characteristic age continues to grow. This can be used to probe field decay
in neutron stars.
 
The problem of magnetic field decay in neutron stars is a long standing one
(see an early discussion in \citealt{jpostriker69NaturePulsarsITheory} and recent
theoretical analysis in \citealt{g2006, caz2004}). Different kinds of
analysis have been used to probe the field evolution. Most often the population synthesis
approach was used to study the whole population of radio pulsars, and
controversial conclusions were reported. 
\cite{b1992} made an important claim that there is no significant field
decay during pulsar lifetime. Recently, \cite{faucher2006} also concluded
that the decay is not necessary to describe the observed population of radio pulsars. 
Oppositely, \cite{g2002} presented arguments in favour of a decaying field. 
\cite{popov2010} presented a model in which several populations of neutron stars
(magnetars, cooling near-by neutron stars, and radio pulsars) have been
explained within the framework of a unique model of magneto-rotational
evolution. However, for ordinary radio pulsars the effect of field decay is
not very pronounced, and so it is difficult to uncover it.
Studies of Be/X-ray binaries have generally confirmed this model (\citealt{cp2012}). 

Magnetic field decay can be highly non-uniform during the lifetime of a neutron
star. Thus analysis of the field evolution over a relatively long time
interval can be, in some sense, misleading. On one hand, it is very
important to put constraints on the very long timescale evolution of the
field. In the near future this can be done on a time scale of billions years, 
for example, if old isolated neutron stars  accreting from the
interstellar medium are discovered (\citealt{kp1997, pp2000, bp2010}). On another
hand, it is useful to analyse field evolution at different -- even relatively
short -- periods of time. In this paper we study the magnetic field evolution of
normal radio pulsars with ages $\sim 10^5$~--~$10^6$~yrs.

We employ our method to estimate the magnetic field decay timescale.
We then test the method with samples of synthetic pulsars generated with a population synthesis code,
and discuss different caveats that may be encountered during the analysis.
Finally, we apply our method to large samples of known pulsars.
Some preliminary results of this study have been reported by \cite{i2014}.

The article is organized as follows. In the next section we describe the main
aspects of the pulsar current analysis and discuss our methodology.  In Section
3 we briefly summarize basic properties of the population synthesis code
which was used to generate synthetic samples of pulsars, and then we apply
these samples to test our method of field decay reconstruction. After that,
in Sec. 3.2,
we study the influence of the source term.
Our main results on field decay in observed radio pulsars are presented
in Section 4.
In Section 5 we discuss uncertainties of the method and, finally, present our  conclusions in the
last section.

\section{Modified pulsar current analysis}

As in the classical pulsar current analysis \citep{pulsar_curr, pb1981}, we
assume that the law describing the time evolution of the magnetic field is
the same for all pulsars and can be written as
 $B(t)=B_0 f(t)$, where $B_0$ is the initial magnetic field (which can be
different for each pulsar) and $f(t)$ is the decay function, which
might be interpreted as a statistical average of the real field evolution
of individual pulsars.  Our goal is to reconstruct 
$f(t)$ from an observational sample of pulsars with measured spin period,
$P$, and period derivative, $\dot P$.  Note that this approach is independent
of the physical mechanism causing the magnetic field evolution.  It simply
provides a purely phenomenological fit to the decay function.

We begin with the following general expression for the magneto-dipole braking
\citep{philippov}:

\begin{equation}
P \dot P = \beta (\kappa_0 + \kappa_1 \sin^2\chi) B^2 ,
\label{field}
\end{equation} 
where $\beta = (\pi^2 R^6)/(I c^3)$, $I$ is the moment of inertia, 
$R$ is the neutron star radius, 
$B\equiv B(t)$ is the magnetic field strength at the magnetic pole,  
$c$ is the speed of light, 
and $\chi$ is the angle between the magnetic axis and the spin axis.
Note that $B(t)$ is a function of time.

The values of the coefficients $\kappa_0$ and $\kappa_1$ determine the
magnetospheric torque.  The most recent 3D simulations for vacuum,
force-free, and resistive magnetospheres \citep{philippov} show that these
coefficients are $\approx 1$ for a variety of magnetospheric models.  The
classical magneto-dipolar radiation formula in vacuum is recovered with
$\kappa_0=0$ and $\kappa_1=2/3$ (\citealt{jpostriker69NaturePulsarsITheory}). 
In this case a neutron star experiences  a very rapid
alignment of the rotation and magnetic axis (see, for example,
\citealt{EliseevaPopov2006}); 
in contradiction with 
observations.  Other alternatives to the magneto-dipole formula (see, for
example, \citealt{gurevich93PhysicsPulsarMagnetosphere, beskin2013} and references
therein) are also similar to Eq.(\ref{field}),
but with a different numerical prefactor or/and different dependence on the angle,
$\chi$.  For our purposes in this paper, a particular choice of the
coefficients is not important.  Hereafter we assume that
$\kappa_0=\kappa_1=1$, and that the evolution of the angle $\chi$ is not
relevant on the timescales we are interested in (this was checked in a
recent study by \citealt{gullon2014}). Therefore, $\sin^2\chi =
\mathrm{const}$,
and for simplicity we assume everywhere below that $\sin^2\chi = 1$.

We treat Eq.(\ref{field}) as a differential equation,
and we combine its solution with the standard definition of 
the spin-down age: $\tau = P/(2 \dot P)$. We then obtain:
\begin{equation}
\label{int_form_final}
\tau(t)=\frac{\beta \int _0 ^t B^2(\tau ') d\tau ' + 0.25 P_0 ^2}{\beta
B^2(t)}.
\end{equation}
We formally average this equation 
over distributions of initial periods and  magnetic fields (see 
Appendix A for details):  
\begin{equation}
\overline{\tau(t)}|_{P_0,B_0} = 
\frac{\int_0^t f^2(\tau ')d\tau '}{f^2(t)}  + \frac{\overline {P^2_0}}{4
\beta \overline {B_0^2}
f^2(t)}=
\label{main_eq}
\end{equation}
$$
= \frac{1}{f^2(t)}\left( \int_0^t f^2(\tau ')d\tau ' +
\frac{\overline{P^2_0}}{4\beta \overline{B_0^2}}\right),
$$
where $\overline{P_0^2}/(4\beta \overline{B_0^2})$ may
be considered as an averaged initial spin-down age. 
This value can be also understood as the median initial spin-down age: half of pulsars 
have their initial spin-down ages smaller than $\overline{\tau_0}$.

Then we differentiate Eq. (\ref{main_eq}) by $t$ and obtain:
\begin{equation}
\label{diff_f_3}
\frac{\dot f(t)}{f(t)}=-\frac{\dot \tau (t)}{2\tau (t)} +
\frac{1}{2\tau(t)}.
\label{eq4}
\end{equation}
In Eq.(4) and below (unless the opposite is directly stated) 
we do not use  overline notation for characteristic ages, as
effectively in our method we always deal, not with the $\tau$ of individual,
pulsars but with some smoothed or average values.
It is remarkable that the form of the differential equation does not
depend on the averaged initial spin-down age.

After we integrate Eq.(\ref{eq4}), we obtain:
\begin{equation}
f(t)={\exp\left(\int_{\tau_\mathrm{min}}^t \frac{dt'}{2\tau (t')}\right)}/
{\sqrt{{\tau(t)}/{\tau_\mathrm{min}}}},
\label{diff_f}
\end{equation}
where the value $\tau_\mathrm{min}$ corresponds to the lower boundary of the range
of characteristic ages that we use in our analysis. 
Thus, the problem is reduced to finding a reasonable approximation to the function $\tau(t)$, from which the
field evolution function $f(t)$ can be recovered by numerical 
integration of Eq. (\ref{diff_f}).  
This can be done using the kinetic equation already used to study the $P$~---~$\dot P$
distribution of radio pulsars \citep{BeskinGurevich1986, Phinney1981, Deshpande1995}.

Let us consider a two-dimensional space 
with  the true age, $t$, as the time coordinate, and 
$\tau$ playing the role of the space coordinate. 
Let $n(\tau, t)$ be the pulsar distribution function in this space.  
This is the number of pulsars with spin-down
age from $\tau$ to $\tau + d\tau$ and true age from $t$ to $t+dt$. We can write the following continuity equation for
the pulsar evolution:
\begin{equation}
\frac{\partial n}{\partial t} +\frac{\partial}{\partial \tau}\left(n\frac{d\tau}{dt}\right)=U-V.
\label{cont_equat}
\end{equation}
Here $U$ and $V$ are source terms describing the rates of birth and death of pulsars 
(latter does not necessary imply some switching-off mechanism;
old pulsars can simply become too faint or too narrow--beamed so we cannot detect them anymore). 
Furthermore, we assume that
during a typical period of a pulsar's activity, 
the whole ensemble of sources is in dynamical equilibrium 
and therefore we may neglect the time
variations of pulsar distributions and search for stationary solutions. 
The second (and the strongest) assumption 
is that both source terms can be neglected in some range 
of characteristic ages $[\tau_\mathrm{min}$, $\tau_\mathrm{max}]$ (see Sec.
2.1), and here Eq. (\ref{cont_equat}) simply reduces to:

\begin{equation}
\frac{\partial}{\partial \tau}\left(n\frac{d\tau}{dt}\right)=0.
\label{simple_form}
\end{equation}

Note that
the distribution of spin-down ages $n(\tau)$ 
can be written as: 
\begin{equation}
n(\tau) = \frac{\Delta N}{\Delta \tau} = \frac{\Delta N}{\Delta{t}} \frac{\Delta t}{\Delta
\tau}. 
\end{equation}
In the limit of infinitesimal intervals and for  a constant birth-rate
(represented by $n_\mathrm{br}$)
the equation above takes the form:

\begin{equation}
n(\tau) =  n_\mathrm{br} \frac{dt}{d\tau}.
\label{ntau}
\end{equation}

Then we integrate this equation
to get the cumulative distribution\footnote{The method to reconstruct the field evolution 
function is realized as a computer code ``Spin Down Ages'' (SDA), available on-line at 
http://www.pulsars.info/decay.html}:

\begin{equation}
N(\tau)\equiv \int_{0}^{\tau} n(\tau',t) d\tau' = n_\mathrm{br} t(\tau).
\label{even_simpler}
\end{equation}

If we assume that the magnetic field remains constant up to some
characteristatic age $\tau_\mathrm{min}$, we obtain $\tau = t +
\overline{\tau_0}$ for $\tau < \tau_\mathrm{min}$.
Therefore: 
\begin{equation}
n_\mathrm{br}  = \frac{N(\tau_\mathrm{min})}{\tau_\mathrm{min}-\overline\tau_0} \approx  \frac{N(\tau_\mathrm{min})}{\tau_\mathrm{min}}.
\label{n_br_est}
\end{equation}

If  $\tau > \tau_\mathrm{min}$ then 
a statistical estimate of the true age of radio pulsars can be defined as:
\begin{equation}
t_\mathrm{stat} (\tau) \equiv \frac{N(\tau)}{n_\mathrm{br}}.
\label{t_stat}
\end{equation}

If we invert this expression and substitute the result into Eq. (\ref{diff_f}) 
to perform numerical integration, we can reconstruct the decay function, $f(t)$.
To do this in a systematic manner, we first introduce a logarithmic 
grid for spin-down ages and 
find the cumulative distribution of $\tau$. 
This is a binned distribution, which is subjected to significant fluctuations. 
It is useful to replace this distribution by a smoothed one applying a linear filter (sliding mean in a window).
This filter is determined by the parameter $r_\mathrm{s}$ which is the size of the window. Explicitly:
\begin{equation}
n'_\mathrm{k} =
\frac{1}{r_\mathrm{s}}\sum_{i=-(r_\mathrm{s}-1)/2}^{(r_\mathrm{s}-1)/2} n_\mathrm{k+i}.
\label{filter}
\end{equation}
Here $n'_\mathrm{k}$ is the number of pulsars in the $k$th bin after
filtering, and $n_\mathrm{k+i}$ ---
the number of pulsars in the $k+i$th bin before filtering. 

Finally, we apply the method only in a relatively narrow range of spin-down
ages $[\tau_\mathrm{min}$, $\tau_\mathrm{max}]$. At large values of $\tau$,
different selections effects can be important, and to get rid of them we
define an upper boundary to the spin-down age. At small values of $\tau$
initial parameters of a pulsar can dominate. We assume that $\tau$ can be
represented as a sum of two values: one related to evolution and another 
to the initial parameters. As initial parameters are unknown we use a
procedure of averaging over them (see Appendix A), and select
$\tau_\mathrm{min}$  in such a way as to minimize the effect of the initial
parameters. 
Details of the choice of $\tau_\mathrm{min}$ and $\tau_\mathrm{max}$ are
given in the next subsection.

\subsection{Determination of $\tau_\mathrm{min}$ and $\tau_\mathrm{max}$}
\label{tau_simple}

The choice of these boundaries  is determined by the necessity to avoid
selection effects.  Our method has two natural limitations, which do not
allow us to apply it to very young or very old pulsars.  First, we assume
that pulsars are born with $\tau < \tau_\mathrm{min}$.  However, 
in reality some objects can have initially $\tau >
\tau_\mathrm{min}$, and for them we cannot distinguish between field decay
and large $\tau_0$ \citep{igoshev2013}.  This is one of the sources of
uncertainty in our approach.
The second assumption is that
there is no selection against older pulsars within the range.  However,
older pulsars are usually weaker and cannot be detected at large distances
from the Sun.  It leads to the leakage of aged pulsars closer to the right
boundary of the range.  Let us discuss both limitations in more details.

To choose the left boundary of the range we want to guarantee for most of
the pulsars in a sample that $\tau_\mathrm{min} $ 
is larger than few$\times \overline{\tau_0}$ (see Eq.\ref{main_eq}).
The definition of the averaged spin-down age determined by Eq. (\ref{main_eq})  includes
the average initial spin-down age.
While the first term in the right hand side in parentheses contains all the field evolution, 
the second one is just some additional constant. 
To estimate this term we use the following values: $P_0=0.3$~s, $B=4\times 10^{12}$~G, and 
$\beta = 1.6\times 10^{-39}$~G$^{-2}$~s:
\begin{equation}
\label{est_dist}
\overline\tau_0 = \frac{\overline {P^2_0}}{4\beta \overline {B_0^2}}
\approx 2.8\times 10^4 \, \mathrm{yrs.}
\end{equation}
So, $\overline \tau |_{B_0, P_0}\sim \tau$ (without additional terms) for
relatively old pulsars with ages significantly larger than the one estimated above.
Consequently, our method may be safely applied to pulsars with  spin-down ages
larger than $\tau_\mathrm{min} \sim 8\times 10^4$ years.
To make an estimate of Eq.(\ref{est_dist}), we choose values such that according to
plots in \cite{popov2012} most of pulsars have $P_0<0.3$~s and
$B_0>4\times 10^{12}$~G, i.e. they are born out of the range under study.

\begin{figure}
\begin{center}
    \includegraphics[width=0.85\columnwidth]{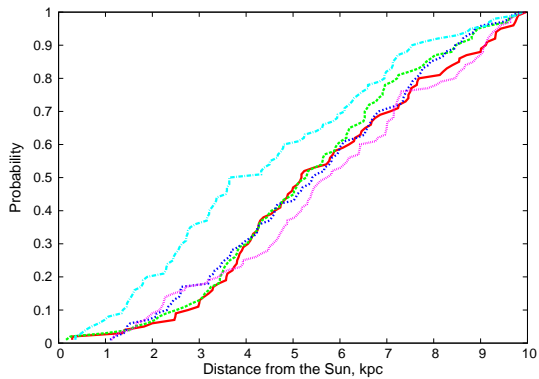}
\end{center}
    \caption{Cumulative distance distributions of pulsars. 
Red solid line  -- $\tau\in[700, 1.2\, \times 10^5]$ years; 
green dashed line -- $\tau\in[1.2\,\times  10^5, 4.5\, \times 10^5]$ years; blue
 short-dashed line -- $\tau\in[4.5\, \times 10^5, 9.6\, \times 10^5]$ years; 
violet dotted line -- $\tau\in[9.6\, \times 10^5, 1.6\, \times 10^6]$ years;
 and light blue dashed and dotted line - $\tau\in[4\, \times 10^6, 6\, \times 10^6]$ years. 
Each age interval contains 100 pulsars.
In this figure we plot all normal pulsars from the ATNF pulsar catalogue 
(\protect\citealt{atnf}), excluding
those in binaries or in globular clusters. (Color on-line.)
}
  \label{obser_select1}
  \end{figure}

To choose the right boundary of the range 
for the real sample (i.e., for a sample of observed pulsars)
we  use the following procedure to probe the leakage of aged pulsars.
Weak pulsars can be hardly ever detected at large distances from the
Sun. Therefore, shapes of radial distribution functions for young and old pulsars
are different because it is not possible to detect weak, aged pulsars with the
same efficiency at all distances, vice versa, till shapes of radial distribution functions for pulsars
of different ages are similar (i.e., untill the difference can be explained by random fluctuations)
the leakage of old pulsars can be neglected. We illustrate this in 
Fig. \ref{obser_select1}.

It is seen that for ages $700$~---~$10^6$~yrs the radial distribution functions
have similar shapes (this is also confirmed by the Kolmogorov-Smirnov test).
However, pulsars with spin-down ages $4\, \times 10^6$~--~$ 6\, \times  10^6$ years
have a radial distribution function with a significantly different shape:
there are more pulsars at small distances than in younger groups. 
This is because some distant, aged pulsars avoid detection due to their weakness, so 
there is a leakage of these sources which can mimic field decay. To avoid
this, we limit our sample to $\tau=10^6$~yrs. 
This value is a bit flexible and potentially can be increased, but to be
conservative we prefer not to do so. 

The similarity of radial distributions might be not sufficient, because even if
these distributions are alike for 
different age groups, some other selection effects which do not influence
the radial distribution can be significant.
Nevertheless, this similarity is a necessary condition because any variation
of the number of pulsars with age due to selection effects, mimic field
evolution.

\section{Population synthesis and tests}

Population synthesis is a numerical method for studying large samples of
evolving objects \citep{pp2007}.  Its most popular variant (which we apply
here) is based on Monte-Carlo procedures which use some initial properties
and evolution laws for individual sources.  Compellingly, selection effects can
also be modelled.  As a result, we create a synthetic sample.  Comparison
between the observed and simulated samples can be done in order to infer
properties of the population.  

\subsection{Tests with synthetic samples}

 The best approach to check the quality of our method is to use a set of
synthetic samples, generated by a robust population synthesis code, 
with several different sets of initial conditions, with
and without field decay, which can more or less successfully reproduce the
real sample of radio pulsars. 
For this purpose we use synthetic samples calculated (and provided to us) by Gull\'on et
al.
Detailed description of their code can be found in \cite{gullon2014} and   
references therein.
Below, we present the most essential details related to the population synthesis code. 

Initial parameters of neutron
stars such as period, magnetic field, position in the Galaxy, and kick
velocity are randomly chosen according to some specified distributions.  The
distributions of initial magnetic fields (in log-scale) and periods are
taken in the form of a Gaussian.  The mean value and standard deviation
vary depending on the model of evolution of the magnetic field
(each model is fitted to reproduce the observed sample of pulsars).  The
considered values can be found in Table \ref{tab:results}.  
The evolution of a pulsar spin
 period is calculated according to \cite{Spitkovsky}
i.e. $\kappa_0 = \kappa_1=1$ in Eq.(\ref{field}) 
The magnetic inclination angle, $\chi$, is uniformly chosen on the sphere, so its direction is isotropic.
Evolution of the magnetic field with time, $B(t)$, 
characterizes each model we use (see Table \ref{tab:results}).
Finally, selection effects are taken into account. 
They determine the fraction of detectable sources among the generated pulsars.
The radio luminosity depends on the spin period and its derivative. 
A popular form for this quantity is used (see, for example,
\citealt{faucher2006}):

\begin{equation}
\log L_\mathrm{rad} = \log [L_0 (P^{-3} \dot P_{-15}) ^ {\alpha}] + L_\mathrm{corr},
\label{lum}
\end{equation}
 where $\dot P_{-15}=\dot P /10^{15}$, 
$L_0=0.18$~mJy kpc$^2$ and $L_\mathrm{corr}$ is chosen
randomly from a Gaussian distribution with zero average and $\sigma=0.8$.
The value of $\alpha$ can vary for different models the magnetic field
evolution (see Table \ref{tab:results}).





\begin{table}
\caption{Results of the SDA code for the synthetic models. $\tau_\mathrm{d}$ corresponds to the timescale
used in the numerical model, while
$\tau_\mathrm{SDA}$ and
$\tau_\mathrm{hist}$ are the ones obtained by applying the SDA code
and a direct fit of $N (\tau)$.}
\label{tab:results}
\begin{tabular}{@{}lccccccccr}
\hline
\hline

Name & $\log{\mu_{B_0}}$	& $\log{\sigma_{B_0}}$ 	& $\mu_{P_0}$  & $\sigma_{P_0}$  & $\alpha$
&$\tau_\mathrm{D}$& $\tau_\mathrm{SDA}$  \\

& [G]	& [G] & [s] & [s] & & [Myr] & [Myr]  \\

\hline

A1	&	$12.60$	&	$0.47$	&	$0.33$	&	$0.23$	&	$0.50$	&	$\infty$	& $\infty$ 	   \\	

A2	&	$12.95$	&	$0.55$	&	$0.30$	&	$0.15$	&	$0.50$	&	$\infty$	& $10$ 	   	  	\\

B1	&	$12.60$	&	$0.47$	&	$0.33$	&	$0.23$	&	$0.50$	&	$0.5$		&$1.00$	   	   \\

B2	&	$12.95$	&	$0.55$	&	$0.30$	&	$0.15$	&	$0.50$	&	$0.5$	 	& $0.690$  	   \\

C1	&	$12.60$	&	$0.47$	&	$0.33$	&	$0.23$	&	$0.50$	&	$1$		& $1.15$   	   \\

C2	&	$12.95$	&	$0.55$	&	$0.30$	&	$0.15$	&	$0.50$	&	$1$		& $0.560$  	   \\

D1	&	$12.60$	&	$0.47$	&	$0.33$	&	$0.23$	&	$0.50$	&	$5$		& $2.00$   	   \\

D2	&	$12.95$	&	$0.55$	&	$0.30$	&	$0.15$	&	$0.50$	&	$5$		& $0.80$   	   \\

E	&	$13.04$	&	$0.55$	&	$0.22$	&	$0.32$	&	$0.44$	&	$\sim 0.8$	& $0.880$  	   \\

\hline
\hline

\end{tabular}
\end{table}

The synthetic samples are created with the following models of evolution of
the magnetic field:

\begin{itemize}

\item Model A. No magnetic field decay: $f(t) = 1$.

\item Models B, C, and D. Exponential decay:  $f(t) = \exp
(-t/\tau_\mathrm{D})$.

\item Model E. A realistic law of field decay based on microphysical
calculations.

\end{itemize}

Models A-D correspond to simplified scenarios, while the last one (E)
 represents a more advanced case. 
 Model E corresponds to a realistic magneto-rotational and thermal evolution
of neutron stars
(see \citealt{vigano2013} and references therein), that was found to
fit well the observational data on radio pulsars (\citealt{gullon2014}).
The Galactic pulsar birth rate in Model E is $\sim2$ neutron stars per century. In
this model the quadratic deviation of the atomic number in the pasta phase
is taken to be $Q_\mathrm{imp} = 25$, as is favoured by a recent study by  \cite{vigano2013}). 
Other parameters are given in Table 1. 
Note, that in Models A-D, the law of
field decay is unique for all pulsars. For Model E this is not the case.
In this model pulsars with different initial parameters follow slightly different paths of
field evolution. 

The initial parameters for all models are listed in Table \ref{tab:results}.
We use two different sets (labeled as $1$ and $2$) for the first four models
(A-D). They are defined by $\log{\mu_{B_0}}$, $\log{\sigma_{B_0}}$,
$\mu_{P_0}$, $\sigma_{P_0}$, $\alpha$ (see Table 1).
For Model E a different set of initial parameters
(that seems to fit better the observational data, see \citealt{vigano2013})
was used.

\begin{figure}
\begin{center}

\begin{minipage}{0.45\linewidth}
\begin{center}
\includegraphics[width=6.1cm]{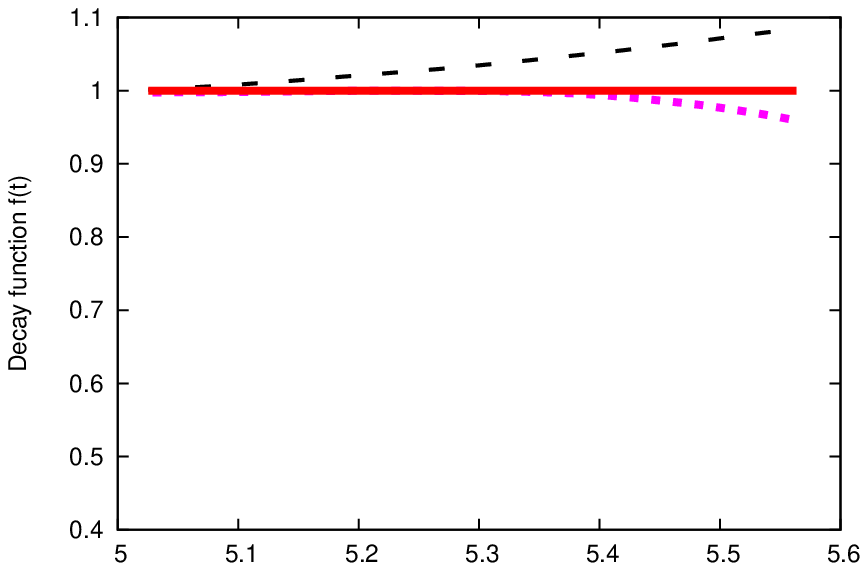}
\includegraphics[width=6.1cm]{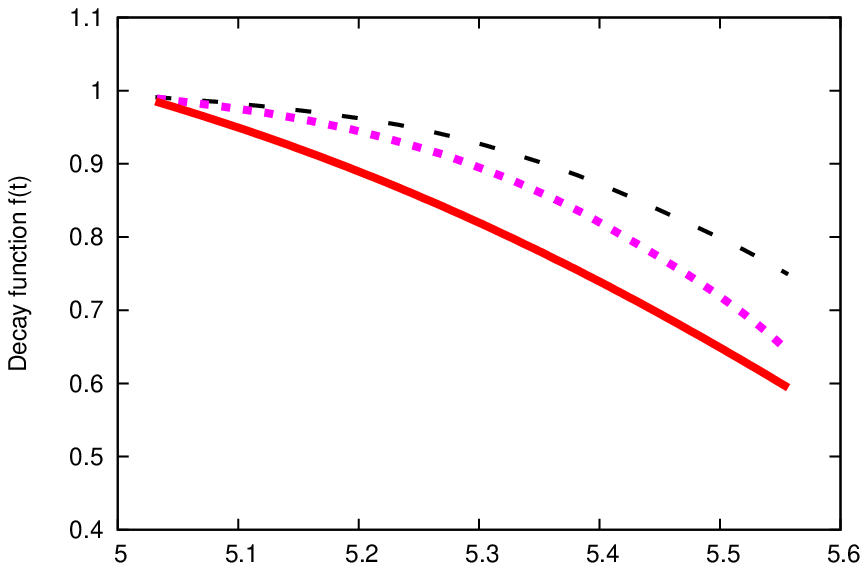}
\includegraphics[width=6.1cm]{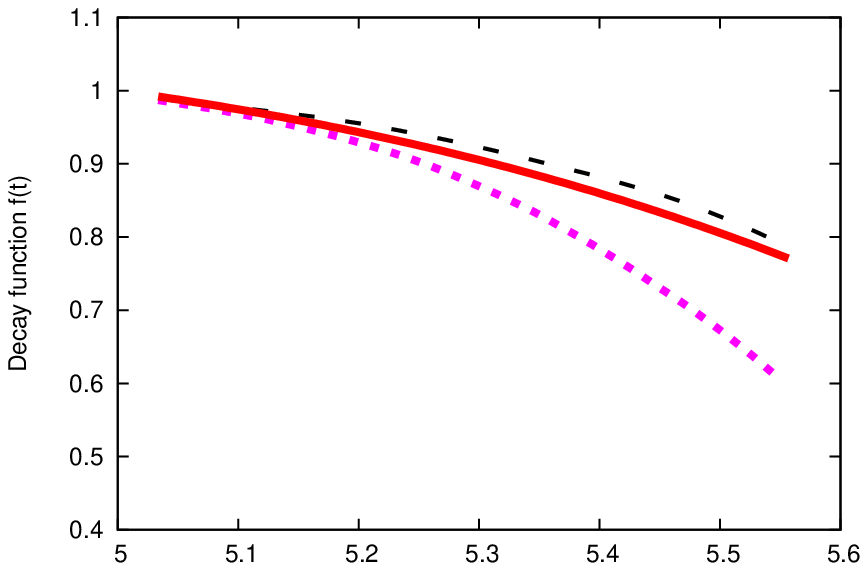}
\includegraphics[width=6.1cm]{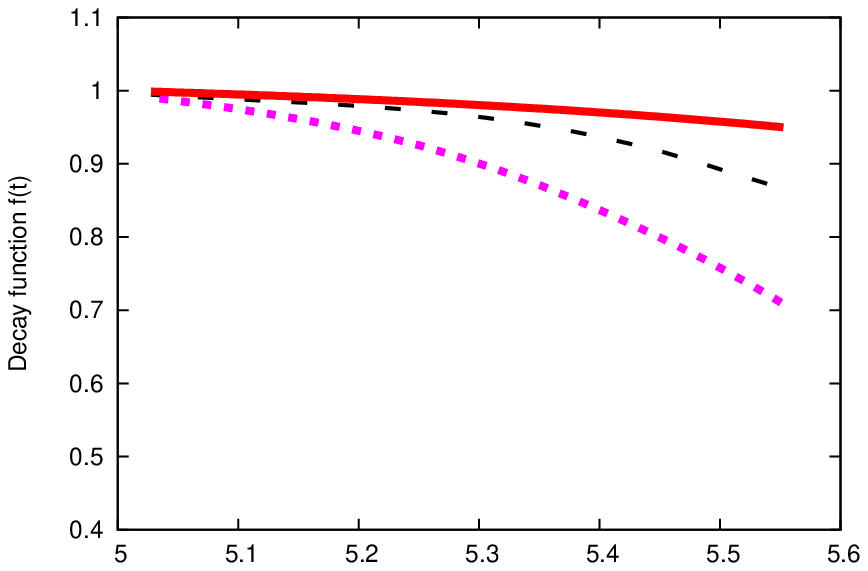}
\includegraphics[width=6.1cm]{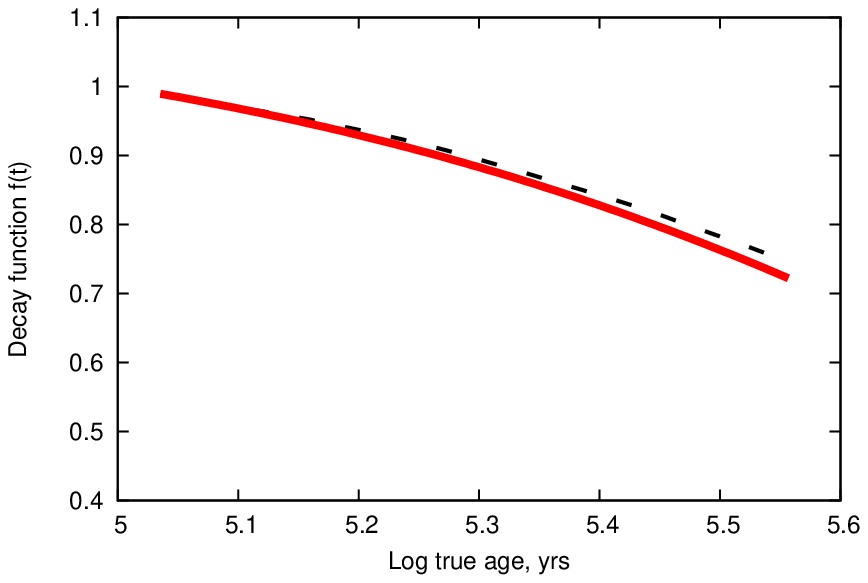}
\end{center}
\end{minipage}
\begin{minipage}{0.45\linewidth}
\begin{center}
\end{center}
\end{minipage}
\end{center}
\caption{The field decay reconstruction for models A-E (from top to bottom). 
In each plot a solid line shows the actual decay law used to generate the
synthetic sample (except the bottom plot where the solid line is an average 
magnetic field decay law, as there the law was not unique for all pulsars).
Dashed lines correspond to  the first set  of initial conditions (models
A1, B1, ...). Dotted lines are plotted for decay reconstruction in which the second set
was used (models A2, B2,....). 
For model E (bottom plot) only one set of parameters have been used, so just one
reconstructed curve is presented, and it is very close to the decay function
used in calculations. 
}
\label{fig:res}
\end{figure}

Since the population synthesis code generates samples with 
pre-defined magnetic field decay law, analysis of the
results and detection of errors are clear.
Errors are divided into random and systematic.
Former ones appear because of the discreetness of the pulsar ensemble; 
while the latter are due to the intrinsic limitations of the method.
For each model we have generated a sample of $10000$ sources.
The results of our tests are presented in Table \ref{tab:results} and Fig. 
\ref{fig:res}.

The following notes can be made:

\begin{itemize}

\item The method is sensitive to the 
magnetic field decay: the obtained timescales systematically increase
for models with slower decay (being maximal for model A).

\item Derived decay timescales in the cases of models B1 and C1 are similar. 
The same is true for samples B2 and C2. However,
the actual values of $\tau_\mathrm{D}$ used to generate each sample in these
pairs  differ by a factor 2.

\item With both methods, when we use the second set of initial parameters (A2, B2, C2, D2)
the derived time scales are always smaller. 

\end{itemize}

\subsection{Influence of the source term}

The last item of the previous subsection
implies that we have some systematics which results in
a more rapid decay if the initial distribution of
characteristic ages is narrower. To analyse this, we perform simple
calculations with a toy-model population synthesis. 

\begin{figure}
\center{\includegraphics[width=84mm]{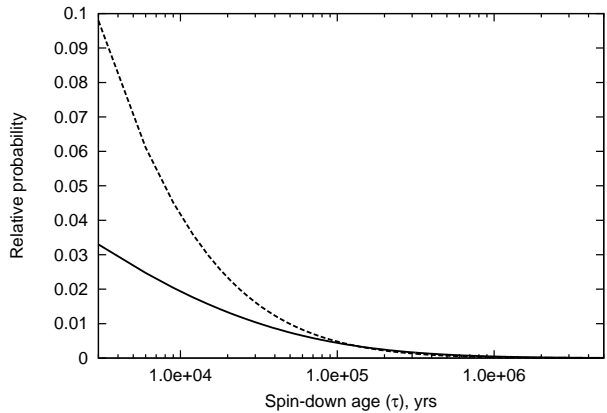}}
\caption{
The PDF of the initial spin-down age. Solid and dashed lines are plotted for the
first and second type of initial condition, correspondingly. Curves are
normalized so that the areas below each of them are equal.
}
\label{init_dist}
\end{figure}

We consider consequent populations of pulsars born with the same initial
distributions with a time step $\Delta t$. In Fig. \ref{init_dist} we plot
distribution of initial characteristic ages for sets 1 (used for models A1,
B1, ...) and 2 (A2, B2, ...). Note, that
for the second set the distribution is narrower (data are normalized in such
a way that the areas below both curves are equal, and the peak of the second
set at low initial characteristic ages is compensated by larger number of
pulsars with initial characteristic ages $>10^5$~yrs in the first set, which is
barely visible in the plot).

\begin{figure}
\center{\includegraphics[width=84mm]{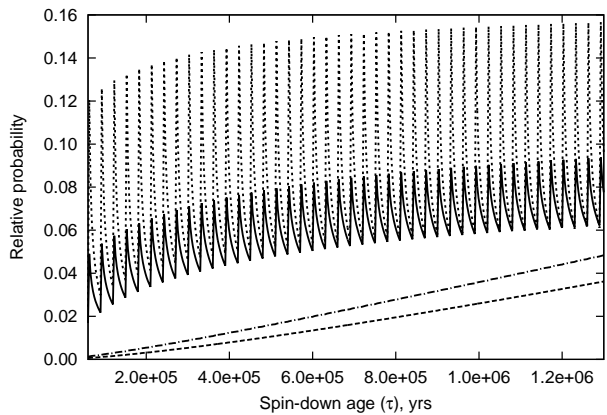}}
\caption{
Results are obtained for constant magnetic field.
Two upper sawtooth curves are PDFs of characteristic ages for several 
sequential generations of pulsars separated by $\Delta t=30000$ years. 
Solid and dashed lines correspond
to the 1st and the 2nd types of initial conditions, respectively. 
Lower dashed and dot-and-dashed
lines are cumulative distribution functions of characteristic ages for the 1st and
the 2nd type of initial conditions, respectively.  For the vertical axis we use
arbitrary units, so curves are shifted respect to each other to show their
behaviour better.
}
\label{peaks1}
\end{figure}

If there is no
field decay, then the distribution of characteristic ages for a
single generation of pulsars is just shifted along the age
axis. 
So the total distribution for all generations would be formed by a
number of peaks separated by $\Delta t$, each of which corresponds to one generation of pulsars.
However, the height of each peaks would be different because, in the total
distribution at each characteristic age there is a contribution from
younger generations. So, peaks which correspond to older generations will
be systematically higher.  Then the cumulative distribution of
characteristic ages is growing faster, and this function would have positive
second derivative (Fig. \ref{peaks1}). This effect explains why we, formally,
obtained a growing magnetic field for model A1 (see Fig.
\ref{fig:res}, top plot).

If the field is decaying then the situation is different. At first, peaks
corresponding to different generations would not be equidistant along the
axis of characteristic ages (our analysis is based on this effect).
Due to field decay older
generations would have enhanced characteristic ages as period derivative of
pulsars is rapidly decreasing (the characteristic age grows faster than the
true age if the field decays: $\Delta \tau > \Delta t$).

\begin{figure}
\center{\includegraphics[width=84mm]{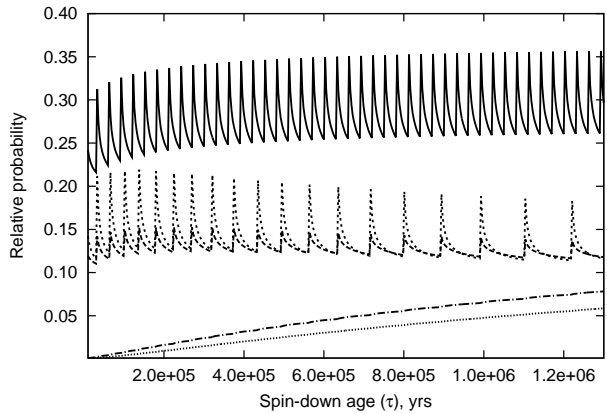}}
\caption{
Three upper sawtooth curves are PDFs of characteristic ages for
 several sequential generations of pulsars separated by $\Delta t=30000$ years.
Solid line on top represents the same PDF as the solid line in Fig. \protect\ref{peaks1}. 
Dotted and dashed sawtooth curves are plotted for the case of decaying magnetic fields for
the 2nd and the 1st type of initial conditions, respectively. Below we
plot two cumulative distributions of characteristic age. The upper of them
is for the 2nd set of initial conditions, and the lower one --- for the 1st.
For the vertical axis we use
arbitrary units, so curves are shifted respect to each other to show their
behaviour better. 
}
\label{peaks2}
\end{figure}

In addition, there is another effect
which we have not included into our analysis. Each
peak is stretching due to decaying field, and so its height decreases (Fig.
\ref{peaks2}, middle curves). As we show below, due to this effect in some cases we
overestimate the rate of field decay.
As the result of growing distances between each sequential generations,
the cumulative distribution has negative second derivative (Fig.
\ref{peaks2}, bottom curves). 

Note that both effects --- growing (due to contributions of younger
generations) and stretching (due to field decay) of the peaks --- 
are 
smaller than the effect of the growing separation between peaks
because the number of pulsars with $\tau_0>\overline{\tau_0}$
is less than a half of the total number (see Appendix B).

\section{Results. Field decay reconstruction}
\label{ATNF_res}








The main goal of this study is to probe the field decay of real radio
pulsars.
We apply our methods to large observed samples of radio pulsars to study
field decay in these objects. As we need to have as large statistics as
possible, as well as uniform samples, we firstly place we study
sources from the ATNF catalogue \citep{atnf}. 
Then we apply our method to the largest uniform subsample of the
ATNF --- to the  Parkes Multibeam and Swinburne surveys (hereafter PMSS) \citep{parkes2001}.
Besides the PMSS, 
the ATNF catalogue  includes Jodrell B \citep{JodrellB}, Green Bank Northern
Hemisphere
survey \citep{GBNHs}, Princeton-NRAO survey \citep{PNS}, Green Bank fast pulsars survey
\citep{fastp}, and 
 other data. 
The PMSS is a significant (major) part of the ATNF pulsar catalogue. This is
the largest relatively uniform sample of radio pulsars.  
We exclude sources not
originally detected in radio surveys (like, magnetars, near-by cooling
neutron stars, etc.)
Also from both samples we exclude millisecond (recycled) pulsars, pulsars in
globular clusters, and in binary systems.
Finally, we use only sources closer than 10 kpc from the Sun. 
In total, we use 1391 objects from the ATNF, and 831 from the PMSS.

\begin{figure}
\begin{center}
\includegraphics[width=8cm]{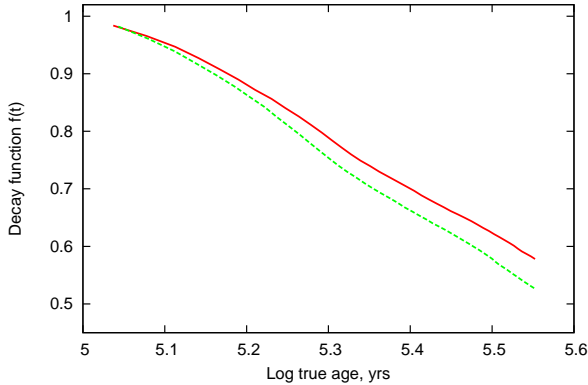}
\end{center}
\caption{Field decay reconstruction for the 
ATNF (dashed line) and PMSS (solid line) samples. 
When fitted with an exponent, the time scale of decay is 0.38 Myrs for the
ATNF sample, and 0.45 Myrs for the PMSS.}
\label{fig:res_real}
\end{figure}

As before for synthetic samples  (see Sec. 3) we reconstruct the magnetic field
decay in the range of true (statistical) ages: $8 \times 10^4 < t 
< 3.5 \times 10^5$~yrs which corresponds to characteristic ages 
$8 \times 10^4 < \tau 
< 10^6$~yrs.  
Results are presented in Fig.~\ref{fig:res_real}. 

The solid line shows the reconstruction for the PMSS data, and the dashed
one for the ATNF. These two curves demonstrate very similar behaviour.
The difference between them is less than 5 per cent.
 This supports the hypothesis that our results are weakly 
dependent on radio fluxes (for the selected range of characteristic
ages and distances). 
In addition, we see that an increase of the number of pulsars by a
factor $\sim 2$ does not influence the results significantly.  
In  Fig.~\ref{fig:res_real} it can be seen visible that the field drops by a factor
$\sim 2$ during the studied period of evolution.
If we fit the derived decay by an exponent, then we obtain:
$\tau_\mathrm{SDA}=0.38$~Myrs for the ATNF sample, and
$\tau_\mathrm{SDA}=0.45$~Myrs for the PMSS sample.
Interpretation of the results is beyond the scope of the paper, but it is
tempting to note that the decay timescale is similar to 
the Hall decay in normal radio pulsars \citep{PonsAguilera}.
Note, that we obtained $f(t)$ just for a limited range of $\tau$. For larger
ages the rate of field decay can be different, and our results cannot be
extrapolated out of the studied range.

\section{Discussion} 

Results of our reconstruction of the magnetic field evolution can be
influenced by several effects related to the assumptions we made.
In this section we briefly discuss them.

In Eq. (\ref{cont_equat}) we neglect both source terms. To 
satisfy this assumption we
choose a lower boundary of the range of characteristic ages;
$\tau_\mathrm{min}$ (see Sec. 2.1). This choice is based on some assumptions
about the initial parameters of neutron stars.  If these assumptions are not
valid, then the decay function is reconstructed with some systematic error.

If the initial distribution of characteristic ages is narrow then we can
overestimate the rate of field decay.
In reality, distributions of $P_0$
and $B_0$ can be wider than we use in our calculations. The fact
that for the most realistic model, E, the reconstructed curves coincide well with the one used
to produce the synthetic samples suggests that this is not a source of
large error for the samples of observed pulsars.


In Appendix B we consider a mathematical model which helps to demonstrate the
dependence of our results on initial distributions. 
This dependence appears to be relatively strong for the cases when
$\tau_\mathrm{min}\sim \overline{\tau_0}$, and weak when
$\tau_\mathrm{min}\gg \overline{\tau_0}$. Note, that if
$\tau_\mathrm{min}\sim \overline{\tau_0}$, then errors in the reconstruction of
$f(t)$ will grow with increasing $t$. Therefore, as in the case of a real sample we
do not know $\overline{\tau_0}$ with good precision, we have an additional reason to limit the
considered range of $t$ from above by only
($t\sim$few$\times\overline{\tau_0}$).


 For the first set of initial conditions (A1, B1, ...) we can estimate that
$\overline{\tau_0} \approx 9\times
10^4$~yrs~$\approx \tau_\mathrm{min}$.
For the second set (A2, B2, ...) we have $\overline{\tau_0} \approx 1.5\times 10^4
$~yrs~$\approx
\tau_\mathrm{min} / 5.5$. Finally,
for the model E the estimate is $\overline{\tau_0} \approx 5\times 10^3
$~yrs~$ \approx
\tau_\mathrm{min}/ 15$. 
This explains why in the case of 
model E, the result of the field evolution reconstruction is in better
agreement with the actual decay, than in the case of A1, B1, ... 

When $\tau_\mathrm{min}\sim \overline{\tau_0}$ the most
important error in the reconstruction of $f(t)$ is related to the
underestimation of the birthrate. This underestimate results in
inadequate reconstruction of $t(\tau)$. Finally, the rate of field decay is
underestimated
(this is an analogue of the effect of summing up in Sec. 3.2).
 
If $\tau_\mathrm{min}\sim 5\overline{\tau_0}$ then
$n_\mathrm{br}\approx C_1$ (see Appendix B1 for description of the
coefficients $C_k$),
and the main influence is due to other terms with coefficients $C_2, C_3$,
etc., which might give a smaller statistical age than the true age.
This happens partly due to the effect that was illustrated in Sec. 3.2 as
the stretching of peaks in the probability density
function (PDF). 

When we apply our method to observed samples (Sec. 4), we use
$\overline{\tau_0}$ from Eq. (\ref{est_dist}) and $\tau_\mathrm{min}=8\times
10^4$~yrs. Then $\tau_\mathrm{min}\sim 3 \overline{\tau_0}$,
and we may overestimate the decay timescale by up to factor $\approx2$.
On the other hand, in Eq. (\ref{est_dist})  we used a rather conservative
estimate of $\overline P_0$. This value is not well known, and if it is
smaller by a factor of a few ($\overline {P_0}=0.1$ instead of $\overline{P_0} = 0.3$ sec,
which would be in congruence with the results
of \citealt{popov2012}). In this case  $\tau_\mathrm{min}\sim
10\overline{\tau_0}$, and our estimates given in Sec.4 are robust.

Still other selection effects can influence the number of observable old pulsars. To
ameliorate this we select an upper limit for the range of characteristic
ages, $\tau_\mathrm{max}$. Still, potentially our results can be
influenced by several effects. Let us discuss them.

On average, radio luminosity of older pulsars can be lower, so some may not be detected.
We studied this possibility by checking the cumulative
distance distributions of pulsars of different ages (Fig. \ref{obser_select1}). 
It seems that our choice of $\tau_\mathrm{max}$ allows us to neglect the
influence of this effect.  

Neutron stars are known to be rapidly moving objects due to large kick
velocities they obtain at birth (\citealt{ll94}). Older pulsars can avoid detection as they
move out of the observable volume (for example, they can move out of the
strip along the Galactic plane where most of the pulsar surveys are
conducted). This would mimic field decay. 
Using numerical integration of pulsar motion in the Galactic 
potential we checked
 how many pulsars with ages $\sim
10^6$~yrs can leave the volume observed by the PMSS. This fraction is about a
few per cent, and we conclude that this effect cannot influence our results
significantly. 

It is known that older pulsars can demonstrate nulling more often than
younger sources \citep{rankin86}. Then some pulsars may not be detected in surveys due to
this effect, and again this can mimic field decay. However, the number
of nulling pulsars is not large. 
Long observational time in each pointing allows modern surveys to detect  even 
pulsars characterized by nulling with a duration of about a few minutes \citep{mclaughlin2003}, 
and we do not expect that this effect can significantly modify our conclusions.

By probing the magnetic field with $P$ and $\dot P$, we always deal with the 
effective field, as the magnetic inclination angle is not known (see Eq.
\ref{field}). If the inclination angle is also evolving, then it is very
difficult to separate real magnetic field evolution from the angle
evolution.  However, recent studies (\citealt{gullon2014}) demonstrate
that the angle evolution in vacuum magnetosphere does not fit the data well --- most
pulsars align too fast. As for plasma-filled magnetospheres the study by
\cite{gullon2014} suggests that the data can be fitted well with the assumption
of non-evolving angle (however, a fit with slightly evolving $\chi$ in the
case of plasma-filled magnetosphere is also
possible, see their Model B1). This allows us to assume that 
the angle evolution can be neglected in the cases we study here.


Finally, in our approach we made an assumption of a unique law of magnetic
field evolution for all neutron stars under study. Without doubt this is
an oversimplification if one studies an ensemble of neutron stars.
For example, extreme magnetars, or central compact objects in supernova remnants can
have very different paths of magnetic field evolution. However, as we are
interested only in normal radio pulsars in a particular range of ages, it seems
reasonable to use in a first approximation the same law of field evolution
for all sources.

\section{Conclusions}
We have proposed and developed a modification of the known pulsar current technique to reconstruct
the magnetic field decay in an ensemble of radio pulsars 
based on spin-down age statistics.

We performed extensive numerical experiments to test our approach,
and these revealed that in many cases, the deduced magnetic field decay law is robust,
although obtained parameters are determined with some uncertainties. 

We performed calculations for normal radio pulsars from the ATNF catalogue, and separately from
the PMSS catalogue, with similar results. This
demonstrates that the method is not particularly sensitive to the number of detected pulsars
used in the analysis, and therefore, to the
certain minimal detectable luminosity.
Also it is found  that the deduced magnetic field decay law
could not be caused by random fluctuations or insufficient sensitivity of modern surveys. 

By analysing pulsars in the ATNF, we find
that in
the range of characteristic ages $8 \times 10^4 < \tau < 10^6$~yrs
(which corresponds to true ages $8 \times 10^4 < t < 3.5 \times 10^5$~yrs)
the effective field decays by a factor $\sim 2$. 
 Taking into account recent results by \cite{gullon2014}, see above, we think that it is
unlikely that all this decay of the effective field can be attributed to the evolution of 
magnetic inclination. We thus conclude that the dipole magnetic
field indeed decays.

The reconstructed decay law is averaged over the entire studied pulsar population
(exact rates of field decay can be different for different subpopulations
among normal radio pulsars). 
The time scale of this decay, when fitted with an exponent,  is about $4\times 10^5$~yrs, which 
is similar to the scale on which  the Hall cascade operates
in normal radio pulsars (for similar range of ages). 
The model with constant fields is shown to be incongruent with the
data. 

This rapid, nearly exponential decay effectively works -- presumably -- only for a
relatively short period of time, and we do not expect that it is still in operation after
$t \sim 10^6$~yrs.

\section*{Acknowledgements}
AI thanks A.F.Kholtygin, V.A.
Urpin, and K.A.Postnov for useful
discussions. We are in debt to Miguel Gull\'on and Jose Pons, who not only
provided samples for tests, but carefully read several versions of the draft
of the paper and made many useful comments and suggestions during our work
on this paper. We thank the unknown referee, who's comments helped to
improve the paper.
We acknowledge David Jones for careful reading of the manuscript and many comments that
helped to improve the text.
SP thanks GGI (Florence) for hospitality during the
workshop ``The Structure and Signals of Neutron Stars, from Birth to Death''.
SP thanks the Dynasty foundation for support of his visit to GGI.
SP was supported by the RFBR grant 12-02-00186.
In the middle of this research AI moved from the Saint Petersburg State University
to the Radboud Universiteit Nijmegen. AI acknowledge Saint-Petersburg State University for a research grant 6.38.18.2014.
AI acknowledges support from the 
Netherlands Research school for Astronomy
(Nederlandse Onderzoekschool voor de Astronomie). 

\bibliography{bibl}

\appendix

\section{Mathematical properties of $\tau$}
\label{aver_app}
\subsection{Averaging}
Nowadays about 1700 isolated, non-millisecond pulsars are known in our Galaxy. 
Every pulsar can be described by a set of parameters.
This set includes magnetic field, spin period, period derivative, radio luminosity, etc. 
Some parameters are physically related, others are independent. 
The spin-down age, $\tau=P/(2 \dot P)$,  is the combination of the two most
important and precisely measured parameters. 

Let $\zeta$ be a parameter of a pulsar (it can be a spin  period, magnetic
field, etc.). Then the distribution function
$\omega(\zeta)$ is defined as the number of pulsars in the interval from $\zeta$ to $\zeta + d\zeta$. 
Let it be a normalized distribution function:
\begin{equation}
\int_a^b \omega(\zeta)d\zeta = 1.
\label{norm}
\end{equation}
Averaging of some other pulsar parameter over this distribution provides an
expectation value for this parameter.  Individual measurements are replaced
by expectation values of the same parameters everywhere in our article.

\subsubsection{Averaging over the initial magnetic field distribution}
\label{aver_b}
The initial magnetic field distribution and initial period distribution 
seem to be independent \citep{popov2012}.
Therefore, we can average over these parameters independently. Let $B_0$ be $\zeta$ in
Eq. (\ref{norm}). It is useful to use the following designation (see also
Eq. \ref{field}): 
\begin{equation}
\label{aver_B}
\overline{\tau(t)}|_{B_0} = \int _{B_1} ^{B_2} \tau(t, B'_0) \omega(B_0')d B_0'.
\end{equation}
Then the expression (\ref{aver_B}) can be rewritten:
\begin{equation}
\label{aver_B_1}
\overline{\tau(t)}|_{B_0} = \overline {\left.\frac{\int _0 ^t \beta B^2(\tau ') d\tau ' + 0.25 P_0
^2}{\beta B^2(t)}\right|}_{B_0}.
\end{equation}
In very young neutron stars (first tens of years of their evolution),
the parameter  $\beta$ can depend on the initial magnetic
field due to star deformations
\citep{ThompsonDincan2000, Ghosh2011, Ghosh2009, jpostriker69NaturePulsarsITheory}.
However, as we study much older objects, we can consider $\beta$
to be independent of $B_0$.
We can rewrite the equation above as:
\begin{equation}
\label{aver_B_2}
\overline{\tau(t)}|_{B_0} = \int _0 ^t \int_{B_1}^{B_2} \frac{B^2(\tau ', B_0)}{B^2(t, B_0)} \omega (B_0) dB_0 d \tau ' +
\frac{P^2_0}{4\beta \overline {B^2}}.
\end{equation}
Here we assume that $B(t) = B_0 f(t)$, where $f(t)$ is a monotonic function. Thereby:
\begin{equation}
\label{aver_B_3}
\overline{\tau(t)}|_{B_0} = \int _0 ^t \int_{B_1}^{B_2} \frac{f^2(\tau ')}{f^2(t)}\omega (B_0) dB_0 d \tau ' +
\frac{P^2_0}{4\beta \overline {B_0^2}f^2(t)}.
\end{equation}
We also assume that the decay function $f(t)$ does not depend on the initial magnetic field
(it is related to the assumption that the function is the same for all
neutron stars). 
The distribution function is normalized
 (\ref{norm}), and therefore, we obtain:
\begin{equation}
\label{aver_B_4}
\overline{\tau(t)}|_{B_0} = \frac{\int_0^t f^2(\tau ')d\tau '}{f^2(t)}  +
\frac{P^2_0}{4\beta \overline {B_0^2} f^2(t)}.
\end{equation}
This is the spin down age with a small disturbance. 

\subsubsection{Averaging over the initial spin period distribution}
\label{aver_p}
Averaging over the initial spin period distribution is similar to the
approach described above.
First, let us consider $P_0$ as $\zeta$. Similarly, we introduce the designation:
\begin{equation}
\label{aver_P}
\overline{\tau(t)}|_{P_0} = \int _{P_1} ^{P_2} \tau(t, P'_0) \omega(P_0')d P_0'.
\end{equation}
It is possible to write:
\begin{equation}
\label{aver_P_1}
\overline{\tau(t)}|_{P_0} = \overline {\left.\frac{\int _0 ^t \beta B^2(\tau ') d\tau ' + 0.25 P_0
^2}{\beta B^2(t)}\right|}_{P_0}.
\end{equation}
Again, in very young neutron stars $\beta$ can be related to $P_0$ due to deformation
of a rapidly rotating object
\citep{jpostriker69NaturePulsarsITheory, Cutler2003}. 
But we can neglect it as we are dealing with older neutron stars.
We write:
\begin{equation}
\label{aver_P_2}
\overline{\tau(t)}|_{P_0} = \int _0 ^t \int_{P_1}^{P_2} \frac{B^2(\tau ')}{B^2(t)} \omega (P_0) dP_0 d \tau ' +
\frac{\overline{P^2_0}}{4\beta B^2}.
\end{equation}
It is assumed that the initial magnetic field and spin period are
independent variables. Therefore, we can write:
\begin{equation}
\label{aver_P_3}
\overline{\tau(t)}|_{P_0} = \frac{\int_0^t f^2(\tau ')d\tau '}{f^2(t)}  + \frac{\overline
{P^2_0}}{4\beta B_0^2 f^2(t)}.
\end{equation}
And again we obtain the spin down age with a small disturbance term.

\subsubsection{Averaging over both distributions}
\label{dist_term}
Now, when Eqs. (\ref{aver_B_4}) and (\ref{aver_P_3}) are known, we average
over both parameters simultaneously:
\begin{equation}
\label{both} 
\overline{\overline{\tau(t)}|_{P_0}}|_{B_0} = \overline{\overline{\tau(t)}|_{B_0}}|_{P_0} := \overline{\tau (t)}|_{P_0,B_0}.
\end{equation}
The result is:
\begin{equation}
\label{both_1}
\overline{\tau(t)}|_{P_0,B_0} = \frac{\int_0^t f^2(\tau ')d\tau '}{f^2(t)}  + \frac{\overline
{P^2_0}}{4\beta \overline {B_0^2} f^2(t)}.
\end{equation}

\section{Analytical description of the algorithm}
In this section we use an analytical approach to 
describe in more detail our method of reconstruction of the decay
function.  This helps to demonstrate more clearly how the
algorithm works without selection effects. To do this we consider several
limiting cases.

\subsection{Distribution of initial spin-down ages}
Let us define a function $\Psi (\tau_0)$ as the probability density
function (PDF) of initial spin down ages $\tau_0$. 
If the PDF of the initial periods is $\Theta(P_0)dP_0$, and the PDF for the initial magnetic fields is
$\Phi(\log B_0)d\log B_0$ (in the following we use notation $b=\log B_0$), 
then the PDF for the initial spin down ages is:
\begin{equation}
\Psi (\tau_0) d\tau_0 = \int _0 ^{\infty} \int _0^{\infty} \delta \left(\tau_0 - \frac{P_0^2}{4\beta B_0^2}  \right) \Theta(P_0) \Phi(b) dP_0 db
d\tau_0.
\end{equation}
This is a sum of probabilities for all initial magnetic fields and periods which
contribute to the spin down age $\tau_0$.
Let us make a substitution $\xi = \tau_0 - P_0^2/(4\beta B_0^2)$. Then we
have:
\begin{equation}
\Psi (\tau_0) d\tau_0 = \int _0 ^{\infty} \Theta (P_0) \int _{-\infty}^{\tau_0} \delta \left(\ \xi \right) \Phi\left(\log \left[ \frac{P_0}{\sqrt{4\beta(\tau_0 - \xi)}}\right]\right) 
\end{equation}
$$
\times\frac{d\xi}{(\tau_0 - \xi)\ln 10} dP_0 d\xi d\tau_0.
$$
Using known properties of the delta function we can simplify this equation:
\begin{equation}
\Psi (\tau_0) d\tau_0 = \int _0^{\infty} \Theta (P_0) \Phi\left(\log \left[ \frac{P_0}{\sqrt{4\beta\tau_0}}\right]\right) \frac{1}{\tau_0\ln 10} dP_0
d\tau_0.
\label{simpl_eq}
\end{equation}
We assume that $\Theta(P_0)dP_0$ and $\Phi(B_0)dB_0$ are Gaussians (for the magnetic field,
the distribution is in a log-scale):
\begin{equation}
\Theta(P_0, \mu_{P_0}, \sigma_{P_0}) dP_0 = \frac{dP_0}{C_\mathrm{norm, 1}} \exp \left( - \frac{(P_0 - \mu_{P_0})^2}{\sigma_{P_0}^2}
\right),
\end{equation}
and
$$
\Phi(\log B_0, \log \mu_{B_0}, \log \sigma_{P_0}) d\log B_0 =
$$
\begin{equation}
\frac{d\log B_0}{C_\mathrm{norm, 2}} \exp \left( - \frac{(\log B_0 - \log \mu_{B_0})^2}{\log \sigma_{P_0}^2}
\right).
\end{equation}
Then Eq. (\ref{simpl_eq}) is expanded to:
$$
\Psi(\tau_0)d\tau_0 = \frac{d\tau_0}{C_\mathrm{norm}} \int_0 ^{\infty} \exp \left( -\frac{(P_0-\mu_{P_0})^2}{\sigma_{P_0}^2} \right)
$$
\begin{equation}
\times \exp  \left( - \frac{\left[\log \left( P_0/\sqrt{4\beta \tau_0} \right) - \log \mu_{B_0}\right]^2}{\log \sigma_{B_0}^2} \right) \frac{1}{\tau_0}   dP_0.
\label{pdf_tau}
\end{equation}
$C_\mathrm{norm}=C_\mathrm{norm, 1} \times C_\mathrm{norm, 2}\times\ln 10$.

Next we define the fraction of pulsars born in different
intervals of the average initial spin-down age:
\begin{equation}
C_1 = \int _0 ^{\overline{\tau_0}} \Psi(\tau_0') d\tau_0' = 0.5,
\end{equation}
and then:
\begin{equation}
C_k = \int _{(k-1)\overline{\tau_0}} ^{k\overline{\tau_0}} \Psi(\tau_0')
d\tau_0'.
\end{equation}

Additionally, to interpret our results for the case of large $\overline{\tau_0}$, it
is useful to introduce $C_{0.5}$:
\begin{equation}
C_{0.5} = \int _0 ^{0.5\overline{\tau_0}} \Psi(\tau_0') d\tau_0'.
\end{equation}

\begin{table}
\caption{Coefficients $C_k$ which describe the shape of the PDF for
initial spin down ages. They are calculated numerically for $\Psi(\tau_0)$ in
the form given in Eq. (\ref{pdf_tau}) for representative values
$\log \sigma_{B_0}=0.5$ and $\sigma_{P_0}=0.2$.}
\label{c_values}
\begin{tabular}{ccccccccccc}
\hline
\hline
$C_{0.5}$ & $C_1$  & $C_2$  & $C_3$  & $C_4$  & $C_5$  & $C_6$  \\
\hline
0.40 & 0.50 & 0.11 & 0.06 & 0.04 & 0.03 & 0.02  \\
\hline
\hline 
$C_7$ & $C_8$  & $C_9$  & $C_{10}$  &  $\sum_{k=11}^{\infty} C_k $ \\
\hline
0.018 & 0.015 & 0.013 & 0.011 & 0.183 \\
\hline
\end{tabular}
\end{table}

Further, we use  two properties of these coefficients:
\begin{equation}
\sum_{k=1}^{\infty} C_k = 1,
\label{sum_c}
\end{equation} 
and
\begin{equation}
C_{k-1} > C_k > C_{k+1}.
\label{small_c}
\end{equation}

Results of numerical integration for the first eleven $C_k$ are listed in the Table \ref{c_values}.
It is worth noting that $C_{0.5}$ is rather large. 

\subsection{Constant field and $n_\mathrm{br}$}
In this subsection we consider the case of constant magnetic field.
Let us suppose that pulsars are born one by one with constant rate: 
each with a time step $\overline {\Delta t}$ after the previous one
($\overline {\Delta t}$ can be also considered as an average expectation time for a pulsar birth).
Then $C_1$ is the fraction of the total number of pulsars in a considered sample  
born with characteristic ages $\tau_0 \le \overline \tau_0$.
For this group of pulsars we can write:
\begin{equation}
\tau(t) \le t + \overline \tau_0.
\end{equation}
For $k$-th group of pulsars we can write:
\begin{equation}
\tau(t) \le t + k\overline \tau_0.
\end{equation}

Therefore, we may use this natural expansion to represent $N(\tau)$:
$$
N(\tau) = C_{0.5} \frac{\tau - 0.5\overline \tau_0}{\overline{\Delta t}} +  (C_1 - C_{0.5}) \frac{\tau - \overline\tau_0}{\overline{\Delta t}} + C_2\frac{\tau - 2\overline\tau_0}{\overline{\Delta t}}
$$
\begin{equation}
+  C_3\frac{\tau - 3\overline\tau_0}{\overline{\Delta t}} + ... = \sum_{k=1}^{\infty} C_k\frac{\tau - k\overline\tau_0}{\overline{\Delta
t}}.
\label{n_def}
\end{equation}
We should terminate this series when the numerator is equal to zero. 
Each term may be split, if necessary, into the sum 
of several terms as  was done in Eq. (\ref{n_def}) for the first term.
It is worth mentioning that each consecutive term is smaller than the previous
one. This is obvious from  
Eq. (\ref{small_c})
and inequality $\tau - k\overline\tau_0 < \tau - (k+1)\overline\tau_0$.

As soon as we estimate the number of pulsars with the spin-down age smaller than
$\tau$, we may introduce a statistical age similar to Eq. (\ref{t_stat}): 
\begin{equation}
T'(\tau) = \sum_{k=1}^{\infty} C_k\frac{\tau - k\overline \tau_0}{n_\mathrm{br}\overline{\Delta
t}}.
\label{time_est}
\end{equation}
In our terminology in this example, 
the true birth-rate is simply $n_\mathrm{br} = 1/\overline{\Delta t}$.
Estimation of this quantity is one step in our method.
When we perform this estimation it we fix some spin down age $\tau_\mathrm{min}$,
and request that $T'(\tau_\mathrm{min}) = \tau_\mathrm{min}$, see Eq.
(\ref{n_br_est}). Then we obtain our estimate of the birthrate:
\begin{equation}
\tilde n_\mathrm{br} = \sum_{k=1}^{\infty} C_k\frac{\tau_\mathrm{min} - k\overline\tau_0}{\tau_\mathrm{min}\overline{\Delta
t}}.
\end{equation}
So, in the limit  $\overline \tau_0 \gg \tau_\mathrm{min}$ we obtain
that indeed $\tilde n_\mathrm{br} = 1/\overline{\Delta t}$, i.e. here our
calculated value is exact. In the case when 
$\tau_\mathrm{min}\approx \overline{\tau_0}$ (which is a bad case),
we obtain $\tilde n_\mathrm{br}=0.2 / \overline{\Delta t}$. 
For better cases  $\tau_\mathrm{min}\approx 3\overline{\tau_0}$
and $\tau_\mathrm{min}\approx 4\overline{\tau_0}$ ,
we obtain $\tilde n_\mathrm{br}=0.37/\overline{\Delta t}$
and $\tilde n_\mathrm{br}=0.445/\overline{\Delta t}$, respectively.  
So, for realistic samples
our estimate of a birthrate 
is between $n_\mathrm{br} / 5 \le \tilde n_\mathrm{br} \le n_\mathrm{br}$.

It is interesting to note that if we estimate 
$\tilde n_\mathrm{br}$ at $\tau_\mathrm{min}\approx 3\overline{\tau_0}$,
and then we restore the statistical age approximately for $5\overline{\tau_0}$, we obtain 
$T'(\tau) = 1.35\tau$ which means that $t>\tau$ and the field formally increases 
(we see it in the case of model A1, Fig. \ref{fig:res}).

\subsection{Decaying field}
In this subsection we consider the case of decaying magnetic field.
Let us again discuss the case when pulsars are born with constant rate with
separation $\overline\Delta t$.
But intervals of spin-down age between two consecutive pulsars during their
evolution are not equal anymore. 
To perform calculations similar to the one presented  in the
previous subsection, 
we need to introduce a new function $G(\tau) = t$ and its inverse $G^{-1}(t) =
\tau$. 
This function allows us to transform a non-uniform grid of spin-down ages $\tau$ to a uniform
grid of $t$.
The function $G^{-1}(t)$ is defined in such a  way that 
when we substitute $G^{-1}(t)$ in the place of $\tau(t)$ into 
Eq. (\ref{diff_f}), then we obtain the exact function $f(t)$.\footnote{Of
course, technically, to define $G$ in such a way we need to know the exact
form of $f(t)$.}
These functions $G(\tau)$ and $G^{-1}(t)$ are the inverse of each other, 
so $G(G^{-1}(t)) = t$ and $G^{-1}(G(\tau)) = \tau$.

To obtain an analogue of 
Eq. (\ref{time_est}) we need to have uniform time intervals in the numerator of this equation. 
To do this we apply our inverse function
to $\tau$ in the numerator of Eq. (\ref{time_est}). This allows us to pass from a
non-uniform grid of $\tau$ to a uniform grid of $t$ 
(the true age does not depend on the magnetic field decay):
\begin{equation}
T'(\tau) = \sum_{k=1}^{\infty} C_k\frac{G(\tau) - k\overline \tau_0}{n_\mathrm{br}\overline{\Delta
t}}.
\label{time_est_decay}
\end{equation}
Let us designate the result of the application of $G$ as $t'_k = G(\tau) - k\overline \tau_0$. 
Then we apply $G^{-1}$ to both sides of Eq. (\ref{time_est_decay}):
\begin{equation}
\tau = G^{-1}\left(\sum_{k=1}^{\infty} C_k\frac{G(\tau) - k\overline \tau_0}{n_\mathrm{br}\overline{\Delta
t}}\right).
\end{equation}
If we substitute this $\tau$ into Eq.(\ref{diff_f}) we obtain $f(t)$:
\begin{equation}
f(T') = f \left( \sum_{k=1}^{\infty} C_k \frac{t'_k}{n_\mathrm{br}\overline{\Delta t}}
\right).
\end{equation}
Basically, in the limit $\tau \gg \overline\tau_0$ 
we have $t'_k \approx t'_1$ and $\tilde n_\mathrm{br} = 1/\overline{\Delta t}$.
Using properties of $C_k$ defined by Eq. (\ref{sum_c}) we obtain that $f(T') =
f(t_1')$.

In the limit $C_1 \approx C_2$ and $\tau_\mathrm{min} 
\approx 2 \overline\tau_0$ we obtain 
$f(T') = f(3t)$. For exponential field decay the relative error 
increases exponentially --- $f(t)/f(3t) = \exp(2t/t_0)$, ---
with timescale $t_0/2$. 
So, in this  case we obtain the decay timescale 
three times smaller than the actual timescale $t_0$.

\end{document}